\documentclass[aps,prb,showpacs,twocolumn,superscriptaddress,groupedaddress]{revtex4-1}

\usepackage{graphicx}  % needed for figures
\usepackage{dcolumn}   % needed for some tables
\usepackage{bm}        % for math
\usepackage{amssymb}   % for math
\usepackage{amsmath}

\usepackage{rotating}

\usepackage{longtable}

\begin{document}

\title{First-principles modeling of the Invar effect in Fe$_{65}$Ni$_{35}$
by the spin-wave method}

\author{A.~V. Ruban}
\affiliation{Department of Materials Science and Engineering, KTH Royal
Institute of Technology, SE-100 44 Stockholm, Sweden}
\affiliation{Materials Center Leoben Forschung GmbH, A-8700 Leoben,
Austria}

\date{\today}
\begin{abstract}
Thermal lattice expansion of the Invar Fe$_{0.65}$Ni$_{0.35}$ alloy is
investigated in first-principles calculations using the spin-wave method,
which is generalized here for the ferromagnetic state with short range order.
It is shown that magnetic short-range order effects make substantial contribution
to the equilibrium lattice constant and cannot be neglected in the
accurate \textit{ab initio} modeling of the thermal expansion in Fe-Ni alloys.
We also demonstrate that at high temperatures, close and above the magnetic
transition, magnetic entropy associated with transverse and longitudinal
spin fluctuations yields a noticeable contribution to the equilibrium 
lattice constant. The obtained theoretical results for the temperature
dependent lattice constant are in semiquantitative agreement with
the experimental data apart from the region close the magnetic transition.
\end{abstract}

\maketitle

\section{Introduction}

Magnetic and structural transformations in Fe are the origin of a
large variety of Fe-based alloys with diverse mechanical and magnetic 
properties, which can be obtained by the proper alloying and 
processing these materials. The fcc Invar FeNi-based alloys
are an example of such a tuning using an alloy composition,
which results in the anomalously low thermal expansion known
as the Invar effect.\cite{guillaume1897}
It exists in a relatively narrow range of
compositions: between ~30 and ~45 at.\% of Ni and most distinct at
35 at. \% of Ni.\cite{at-wt} It can be made even more pronounced
using other alloying schemes.\cite{superInvar}  

The fact that the Invar effect (as well as the Elinvar effect)
is somehow related to magnetism was understood by Guillaume
a century ago who started his Noble lecture\cite{guill_nl} from a
description of magnetic properties, namely the magnetic transition
temperatures of Fe-Ni alloys, although at that time, there was no
sensible theory of magnetism in solids.
Only in 1963, Weiss made that connection clear in his well-known
2$\gamma$-state model\cite{weiss63} arguing that the usual thermal
lattice expansion due to the lattice anharmonicity was compensated in
the Fe-Ni alloys by a temperature induced "electronic" transition
of the states with a higher moment and a large volume to the high
temperature states having a lower magnetic moment and volume.

What was, however, confusing in the Weiss model is
the identification of the high-temperature "$\gamma$-phase" as
antiferromagnetic.  Thirty years later, the low volume and
low magnetic moment state was discovered in a number of
\textit{ferromagnetic} first-principles
calculations.\cite{moruzzi90,mohn91,entel93,schroter95,entel97,podgorny92,peperhoff01}
The connection to the Invar effect due to thermal spin excitations
then was accomplished using a phenomenological Ginzburg-Landau model.

Unfortunately, these calculations and models have little to do
with the finite temperature magnetic state in the real Invar Fe-Ni alloys,
which is ferromagnetic with a certain degree of randomness leading
to the lowering of the magnetization with temperature.
The origin of the problems with these earlier \textit{ab initio}
calculations is the use of the local spin density approximation, which
is as well-known nowadays, significantly underestimate the equilibrium 
volume of 3d-metals and their alloys. So, the theoretical
equilibrium volume was too small and close to the so-called
low-spin -- high-spin transition, which was not actually the
case of the real Fe$_{65}$Ni$_{35}$ alloys.

The first adequate semi-empirical theory of the Invar effect
was developed by Kakehashi\cite{kakehashi81,kakehashi89}
who explained it by a reduction of the local moments of Fe due to
temperature induced magnetic disorder. Such a reduction of the
magnetic moment in the magnetically disordered state compared
to the ground state ferromagnetic one was for the first time
confirmed in the first-principles calculations by
Johnson {\it et al.} \cite{johnson90,johnson97}
using the disordered local moment (DLM) model\cite{gyorffy85}
for the paramagnetic state and later by Akai and
Dederichs.\cite{akai93}

Thus the qualitative picture of the Invar effect is related to
the reduction of the equilibrium 0 K ground state volume of
the finite-temperature ferromagnetic phase due to
the increasing randomness of the magnetic configuration. If one
assumes that the thermal expansion due to the lattice anharmonicity
is fixed, i.e. does not depend on temperature, the description of
the Invar effect will be reduced to the finding the 0 K
equilibrium volume of the alloy in the magnetic state with the
reduced magnetization, which corresponds to the given temperature.

This simple model was used in the first-principles calculations by
Crisan \textit{et al.}\cite{crisan02}
who almost quantitatively reproduced the experimental thermal
expansion coefficient of Fe$_{65}$Ni$_{35}$ between 0 and 1000 K.
Although some of the details of this modeling are
questionable,\cite{ruban07} nevertheless, this was the first
\textit{ab initio} investigation, which reproduced the Invar effect.

Similar first-principles-based modeling of the Invar effect,
using 0 K total energies of alloys with different spin 
configurations representing the finite-temperature magnetic state
of the Invar alloy, was successfully applied by
Khmelevskyi {\it et al.} to a number of different Invar
systems,\cite{khmelevskyi03,khmelevskyi04a,khmelevskyi04b,khmelevskyi05,ruban07}
A more elaborate approach was adopted by Liot and co-authors \cite{liot09,liot12,liot14} who actually calculated the finite
temperature lattice constant and thermal expansion coefficient
of some Invar alloys using the Debye-Gr\"uneisen model.\cite{moruzzi88}

It is obvious, that the above mentioned computational schemes are approximate
in many details. The most important one, especially in the case of
Fe-Ni Invar alloys, is a quite approximate description of the finite
temperature magnetic state in the DFT based calculations. Unfortunately,
going beyond DFT with the proper quantitative modeling of the Invar
effect is unrealistic at the present time. 
Nevertheless, until the proper methods and computers are available,
one can still try to improve the existing DFT-based models.

In particular, in this work, a more elaborate model of the thermal
expansion in the Invar Fe$_{65}$Ni$_{35}$ alloy is developed. It
1) includes a consideration of the magnetic short-range order (MSRO)
effects both below and above the Curie temperature and 2) takes into
consideration magnetic entropy related to the temperature induced spin
fluctuations. The MSRO effects are incorporated using the spin-wave
method (SWM),\cite{ruban12} which is generalized here for
the ferromagnetic state with arbitrary magnetization.

Heisenberg Monte Carlo simulations with magnetic exchange 
interaction parameters obtained in first-principles calculations,
are done to determine the magnetic state characterized by the
corresponding magnetic short- and long-range order (LRO) for
every temperature. This information is used in the SWM calculations
to get the total energy of the system in the given magnetic state
and parameters of the Debye-Gr\"uneisen model yielding finally the
lattice constant\cite{moruzzi88,korzhavyi94} at the corresponding 
temperature.

\section{Spin-wave method for systems with magnetic long-
and short-range order} \label{sec:SWM}

The SWM is based on the assumption that the
magnetic energy of a system is given by Heisenberg Hamiltonian

\begin{equation}\label{eq:H}
H = - \sum_p \sum_{i,j \in p }
J_p \mathbf{e}_{i} \mathbf{e}_{j} ,
\end{equation}
where $J_p$ are the magnetic exchange interaction parameters,
which do not depend on the magnetic state; $p$ is the coordination
shell and $\mathbf{e}_{i}$ is the direction of the magnetic moment at
site $i$.
  
For such a Hamiltonian, the magnetic configuration is uniquely
identified by spin-spin correlation functions

\begin{equation}
\xi_p \equiv <\mathbf{e}_{i} \mathbf{e}_{j}>_p =
\frac{1}{N z_p}\sum_{i,j \in p} \mathbf{e}_{i} \mathbf{e}_{j} ,
\end{equation}
where $N$ is the number of atoms and $z_p$ is the coordination number.
The magnetic energy of a system presented by Hamiltonian (\ref{eq:H})
then can be determined as

\begin{equation}\label{eq:E_JR}
E = - \sum_p J_p z_p \xi_p =
 - \sum_{\mathbf{R}} J(\mathbf{R}) \xi(\mathbf{R}) .
\end{equation}

Since, in the ideal paramagnetic (IPM) state $\xi(\mathbf{R})=0$,
and therefore (\ref{eq:E_JR}) is just the energy of the magnetic
long- and short-range order.

Using the following definition of the Fourier transform of 
the spin-spin correlation function

\begin{equation} \label{eq:xi_q}
\xi(\mathbf{q}) = \sum_{\mathbf{R}} \xi(\mathbf{R}) e^{i \mathbf{qR}} ,
\end{equation}
magnetic energy (\ref{eq:E_JR}) can be determined in another form:

\begin{equation} \label{eq:E_Jq}
E = \frac{1}{\Omega_{BZ}} \int_{BZ} d\mathbf{q}
J(\mathbf{q})\xi(\mathbf{q}),
\end{equation}
where $J(\mathbf{q})$ is the Fourier transform of the 
magnetic exchange interactions (see, for instance,
Ref. \onlinecite{halilov98}):

\begin{equation} \label{eq:J_q}
J(\mathbf{q}) = -\sum_{\mathbf{R}} J(\mathbf{R}) e^{i \mathbf{qR}} ,
\end{equation}
which is, up to an additive
constant, just the energy of the planar spin-spiral (PSS) with
wave vector $\mathbf{q}$. 

The spin-spin correlation function of such a PSS
is\cite{sandratskii98}

\begin{equation} \label{eq:xi_qR}
\xi_{\mathbf{q}}(\mathbf{R}) = \cos(\mathbf{qR}) ,
\end{equation}
and the equally weighted superposition of all the PSS
with different wave vectors $\mathbf{q}$ yields the IPM since

\begin{equation} \label{eq:xi_R}
\frac{1}{\Omega_{BZ}} \int_{BZ} d\mathbf{q} \xi_{\mathbf{q}}(\mathbf{R})
=\xi(\mathbf{R})  =  0
\end{equation}
for all $\mathbf{R}$ except for $\mathbf{R}=0$, which
$\xi(\mathbf{R}=0) = 1$.

The Fourier transform of $\xi_{\mathbf{q}}(\mathbf{R})$ for some
specific $\mathbf{q}'$ is the Dirac $\delta$-function,
$\xi_q(\mathbf{q^{\prime}}) = \delta(\mathbf{q}-\mathbf{q}^{\prime})$,
and therefore the following important normalization condition
holds:

\begin{equation} \label{eq:norm}
\frac{1}{\Omega_{BZ}^2}\int_{BZ} d\mathbf{q}
\int_{BZ} d\mathbf{q}^{\prime}\xi_{\mathbf{q}}(\mathbf{q}^{\prime}) = 1.
\end{equation} 

This means that the PSS form a complete and orthogonal basis with
eigenvalues $E(\mathbf{q})$, which are in general the total energies
of a system within some general Hamiltonian, in the the PSS 
magnetic configuration with wave vector $\mathbf{q}$. In the case of
Heisenberg Hamiltonain (\ref{eq:H}), $E(\mathbf{q}) = J(\mathbf{q})$.
If $E(\mathbf{q})$ is obtained in first-principles calculations
and the magnetic energy of the system can be described by 
Hamiltonian (\ref{eq:H}), $E(\mathbf{q}) = J(\mathbf{q}) + E_0$,
where, as will be shown below, $E_0$ is the total energy of the IPM
obtained in the same first-principles calculations. 

The energy of the IPM state within the first-principles formalism,
can be found using the fact that its real space spin-spin correlation
functions are given by the equal weighted superposition of all the PSS
in the reciprocal space (\ref{eq:xi_R}).
Then one can use (\ref{eq:E_Jq}) by substituting $E(\mathbf{q})$
instead of $J(\mathbf{q})$, which defines the energy of specific magnetic
configuration given by a set of spin-spin correlation functions,
$\xi(\mathbf{R})$ or equivalently $\xi(\mathbf{q})$, so that

\begin{eqnarray} \label{eq:E_IPM}
E^{\rm IPM} &=& \frac{1}{\Omega_{BZ}^2}
\int_{BZ} d\mathbf{q} E(\mathbf{q}) \int_{BZ} d\mathbf{q}^{\prime}
\xi_{\mathbf{q}}(\mathbf{q}^{\prime}) \\ \nonumber
&=& \frac{1}{\Omega_{BZ}} \int_{BZ}d\mathbf{q}
\ E(\mathbf{q}) .
\end{eqnarray}

If one uses $J(\mathbf{q})$ instead of $E(\mathbf{q})$ in (\ref{eq:E_IPM}),
the last integral is formally equal to $J(\mathbf{R}=0)$, which is routinely
accepted to be zero in (\ref{eq:H}). Thus $E^{\rm IPM} = E_0$ and it indeed
connects the Fourier transform of the magnetic exchange interactions of
Hamiltonian (\ref{eq:H}) $J(\mathbf{q})$  and $E(\mathbf{q})$ from
first-principles calculations.

The presence of the MSRO leads to a deviation of $\xi^{\rm SRO}(\mathbf{q})$,
or simply $\xi(\mathbf{q})$, from the equal weighted
distribution of the PSS in the reciprocal space. However, its does not
violate the normalization of the expansion in terms of the PSS since

\begin{equation} \label{eq:norm}
\frac{1}{\Omega_{BZ}}\int_{BZ} d\mathbf{q}
\xi(\mathbf{q}) = 0 ,
\end{equation} 
and thus the energy of the MSRO is

\begin{equation} \label{eq:Delta_E_MSRO}
\Delta E^{\rm SRO} = \frac{1}{\Omega_{BZ}} \int_{BZ}d\mathbf{q}
\ E(\mathbf{q}) \xi(\mathbf{q}) .
\end{equation}
Then the total energy is given by the sum of the energy of the IPM state
and MSRO:

\begin{equation} \label{eq:E_MSRO}
E = \frac{1}{\Omega_{BZ}} \int_{BZ}d\mathbf{q}
\ E(\mathbf{q}) \left(1 + \xi(\mathbf{q}) \right) ,
\end{equation} 
The later expression defines the energy of paramagnetic
state with MSRO and can be used in first-principles calculations.

The outlined above formalism should be modified in the
presence of magnetic long-range order. Here, it is done for
the ferromagnetic state. In this case, spin-spin correlation
functions $\xi(\mathbf{R})$, and respectively $\xi(\mathbf{q})$, 
can be divided below the Curie temperature into two contributions:
from the long-range order, $s_0$, and short-range order,
$\xi^{\rm SRO}(\mathbf{R})$, defined in the following way:

\begin{eqnarray}
&&s_0 = \lim_{R \to \infty} \xi(\mathbf{R}) ; \\
&&\xi^{\rm SRO}(\mathbf{R}) = 
\xi(\mathbf{R}) - s_0 .
\end{eqnarray}
It is clear that $s_0$ is the long-range order parameter, or
the reduced magnetization in the case of a ferromagnet.

The Fourier transform of $\xi(\mathbf{R})$ will contain
then also two contributions: from the long-range order, which
is $s_0\delta(\mathbf{q}-0)$ and the contribution from
the short-range order part, $\xi^{\rm SRO}(\mathbf{q})$, which
is the Fourier transform of $\xi^{\rm SRO}(\mathbf{R})$.
The later does not contribute to the total normalization, and
therefore, in order to have the proper normalization, one should
add the completely random background compensating the missing
part from the long-range order contribution, i.e. normalized
by $1 - s_0$. Thus, the total energy of the FM state
with reduced magnetization $s_0$ is

\begin{eqnarray} \label{eq:E_fm}
E &=& s_0 E(\mathbf{q}=0) + \\ \nonumber
&&\frac{1}{\Omega_{BZ}} \int_{BZ}d\mathbf{q}
\ E(\mathbf{q})\left[1 -s_0 +
\xi^{\rm SRO}(\mathbf{q})\right] .
\end{eqnarray}
The first term is just the energy of the completely
ordered FM state, while the second one is the contribution
from the randomly oriented magnetic moments (with some
MSRO), which reduces the magnetization of system from
its completely ordered value.

If one neglects the SRO effects in the second term, i.e.
if $\xi^{\rm SRO}(\mathbf{q})=0$, Eq. (\ref{eq:E_fm}) defines
the so-called partial disordered local moment (PDLM) model, which
is used for the modeling of a ferromagnetic
state at finite temperatures within the coherent potential
approximation (CPA) calculations.

%%%%%%%%

\section{Details of \textit{ab initio} calculations}

According to the existing experimental data\cite{robertson99,tsunoda08}
as well as the results of first-principles modeling,\cite{ruban07}
Fe-Ni Invar alloys are random, without noticeable atomic short-range
order. The easiest way to determine the electronic structure and 
the total energy of such alloys is to use the CPA,\cite{CPA} 
which works very well for this systems as has been demonstrated
previously\cite{ruban07} and also confirmed in this work.
The CPA calculations have been done in the framework of the
Green's function EMTO method.\cite{andersen94,andersen98,vitos01}
In particular, the Lyngby version of the EMTO code has been used,
which allows non-collinear
and spin-spiral calculations as well as the correct treatment of the
screened Coulomb interactions within the single site approximation.\cite{ruban02} 
In the latter case, the on-site screened electrostatic potential,
$V^i_{scr}$, and energy, $E^i_{scr}$\cite{ruban02}, are 

\begin{eqnarray}  \label{eq:scr_0}
v_{scr}^i &=& -e^2 \alpha_{scr} \frac{q_i}{S} \\
E^i_{scr} &=& -e^2 \frac{1}{2} \alpha_{scr} \beta_{scr} \frac{q_i^2}{S} .
\end{eqnarray}
Here $q_i$ is the net charge of the atomic sphere of the $i$th alloy
component, $S$ the Wigner-Seitz (WS) radius, and $\alpha_{scr}$ and
$\beta_{scr}$ the on-site screening constants. The screening constants
of a random Fe$_{65}$Ni$_{35}$ alloy in the FM and DLM states have been
obtained in the 560-atom supercell locally-self consistent Green's function
(LSGF)\cite{abrikos97} calculations using the ELSGF method\cite{peil12}
including the first two coordination shells in the local interaction
zone.\cite{abrikos97}
The screening constants are found to be very little dependent on the
magnetic state and the lattice constant, and $\alpha_{scr}=$ 0.8 and
$\beta_{scr}=$1.14, have been then used in all the EMTO-CPA 
calculations. 

The basis functions in all the EMTO calculations have been expanded up to
$l_{max}= $3. We have also taken into consideration the multipole moment
contributions to the electrostatic energy. The summation over multipole
moments for the electrostatic part of the one-electron potential and
total energy have been carried out up to $l_{max}=$6. The integration
over the irreducible part of the Brillouin zone has been performed using
the 36$\times$36$\times$36 Monkhorst-Pack grid.\cite{monkhorst72}

The self-consistent calculations have been done within the local
density approximation (LDA),\cite{LDA92}
while the total energy has been obtained using the PBE generalized
gradient functional (GGA).\cite{pbe96}
The point is that the LDA self-consistency does not affect the final
GGA total energy, but it affects the magnitude of the magnetic moment,
which is usually slightly overestimated in the PBE-GGA compared with
the LDA one. 

In spite of the existing theoretical investigation of local
lattice relaxations in Fe$_{65}$Ni$_{35}$ in the FM
state,\cite{liot09rel} nothing is known about their effect on
the thermal lattice expansion and the Invar effect, in particular.
Therefore, in order to estimate the role of local lattice relaxations
in the energetics of Fe-Ni Invar alloys, we have calculated
the equilibrium lattice constants and total energies of a
64-atom supercell (4$\times$4$\times$4) modeling Fe$_{62.5}$Ni$_{37.5}$
random alloy\cite{scell} with relaxed (locally) and unrelaxed
atomic positions. 

The PBE-GGA calculations have been done by the projector augmented
wave (PAW) method\cite{blochl94,kresse99} as
implemented in the Vienna {\it ab initio} simulation
package (VASP) code.\cite{kresse93,kresse94,kresse96} 
We find that the lattice constant of the supercell with relaxed atomic
positions is only 0.0006 \AA\ smaller than that of the unrelaxed one.
The relaxation energy is also small: 7 meV/atom. This means
that local lattice relaxations can hardly have any effect on the
Invar effect, which is in fact expected from the relatively small
size difference of Fe and Ni.

\section{Models of finite temperature magnetism in
F\lowercase{e}-N\lowercase{i} alloys}

Magnetism plays crucial role in the Invar effect, which exists
in Fe-Ni alloys in the FM state. However, the paramagnetic state
is not less important, since it determines the type of thermal
magnetic excitations both above and below the Curie temperature.
In all the previous \textit{ab initio} modeling of the Invar effect
in FeNi, it has been tacitly assumed that the main thermal magnetic
excitations is the thermal disorder of the orientation of the local
magnetic moments, which magnitude remains unchanged, or
transverse spin fluctuations (TSF).

However, the magnitude of the local magnetic moments on Fe and Ni
in Fe$_{65}$Ni$_{35}$ is quite sensitive to the magnetic state,
and, for instance, the magnetic moment on Ni vanishes in the
DLM representation of the paramagnetic state. This does not correspond
to what really happens in the paramagnetic state at finite temperature,
where non-zero magnetic moment exists due to an entropic effect of
thermally induced longitudinal spin fluctuations (LSF).
As will be demonstrated below, the LSF play important role in
Fe$_{65}$Ni$_{35}$ and therefore, in this section, we introduce
a simplified model of the LSF, which will be used further both in the 
calculations of magnetic exchange interactions and in the modeling
of the thermal lattice expansion.
 
\subsection{Longitudinal spin fluctuations in Fe$_{65}$Ni$_{35}$}

Unfortunately, accurate first-principles modeling of the
LSF is too cumbersome for the case of random alloys. An approximate
DFT-based model developed in Ref. \onlinecite{ruban07_lsf} is also
quite complicated for Fe-Ni alloys due to pronounced local chemical
environment effects,\cite{ruban05} which require an additional consideration
of different possible local atomic configurations. Therefore, a
simple model is used here for just for a qualitative estimate of
LSF, which has been previously introduced in Ref. \onlinecite{ruban13}.

Here, we consider LSF in the paramagnetic state given by the DLM model
assuming that they are adiabatically connected with a particular
electronic structure, which results in specific magnitudes of the local
magnetic moments of alloy components. In this case, the local magnetic
moment of specific alloy component, $i$, at a given temperature T can
be determined in the single-site mean-field approximation
as\cite{ruban13,sandratskii08}

\begin{equation} \label{eq:m_lsf}
<m_i> = \frac{\int m^3 \exp\left[-\beta E^{\rm LSF}_i(m)\right] dm}
{\int m^2 \exp\left[-\beta E^{\rm LSF}_i(m)\right] dm} , 
\end{equation}
where, $\beta = 1/T$ and $E^{\rm LSF}_i(m)$ is the so-called LSF energy
of the $i$th component, which is the total energy of an alloy per atom of 
this alloy component. It is determined here for each alloy
component in the constrained DLM-CPA calculations by fixing the local
magnetic moment of alloy component $i$, while allowing the magnetic
moment of the other alloy component to relax to the corresponding
self-consistent magnitude.  

\begin{figure}[tbp]
\includegraphics[height=8.0cm]{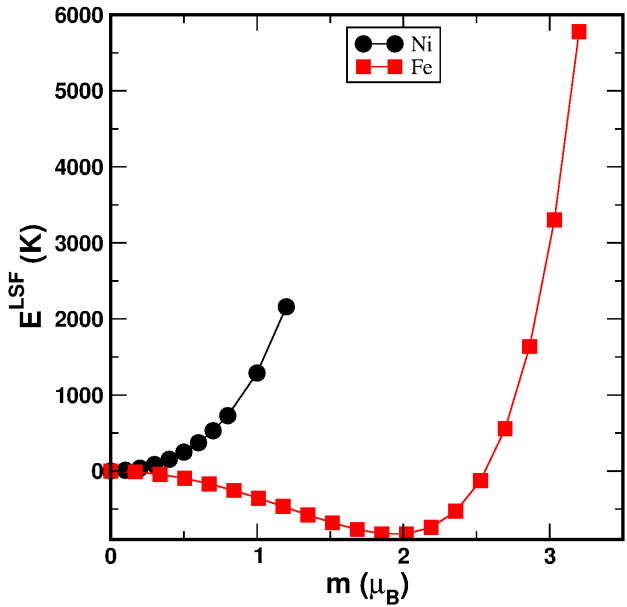}
\caption[3]{(Color online) LSF energy of Ni and Fe in 
random Fe$_{65}$Ni$_{35}$ at 1000 K.}
\label{fig:E_lsf}
\end{figure}

In Fig. \ref{fig:E_lsf}, the LSF energy of Ni and Fe in
Fe$_{65}$Ni$_{35}$ is shown for the lattice constant
of 3.59 \AA. The calculations are done at 1000 K and the LSF
energy contains also the entropy contribution from the
one-electron excitations and in the case of the LSF energy of
Ni  also contribution from the magnetic entropy of Fe taking in
the usual form of the magnetic entropy of paramagnetic gas:

\begin{equation} \label{eq:S_TSF}
S^{\rm TSF} = \ln(m^{\rm Fe} + 1) ,
\end{equation}
where $m^{\rm Fe}$ is the magnetic moment of Fe in the DLM-CPA
calculations. The latter, in fact, corresponds to the magnetic
entropy of the TSF.  Thus, the LSF energy of Ni contains also
the entropy contribution from the one-electron excitations on
both components and magnetic entropy (\ref{eq:S_TSF}) on Fe.

This approach, however, does not work for the LSF energy of Fe,
since the magnetic moment of Ni vanishes in the DLM-CPA calculations
and therefore one should consider LSF on Ni at the given
temperature. As has been demonstrated in Ref. \onlinecite{ruban13}, 
if the LSF energy has a quadratic form (as a function of magnetic
moment), the LSF entropy is 

\begin{equation} \label{eq:S_lsf}
S^{\rm LSF} = 3 \ln(<m_i>) ,
\end{equation}
where $<m_i>$ is the average magnitude of magnetic moment
of the $i$th alloy component.
This expression is valid in the high-temperature (classical)
limit. In the single site approximation, one can simply
substitute $<m_i>$ by $m_i$, and $m_i$ can be found from
the free energy minimization of the alloy.

The latter is however a quite tedious procedure.
Fortunately, this task can be substantially simplified in
the DFT self-consistent calculations if one notice that
such a minimization results in the appearance of an 
additional contributions to the one-electron potential,
which can be obtained from the corresponding functional
derivatives of (\ref{eq:S_lsf}) with respect to spin-up
and spin-down densities.\cite{korzhavyi} Then the magnetic
moment induced by the LSF at a given temperature is
just the result of the corresponding DFT self-consistent
calculations, which simplifies enormously the whole
calculation procedure. 

Therefore, in order to account for the (qualitatively)
correct magnetic state of Ni during the LSF energy calculations
of Fe, the LSF on Ni were taken into consideration using
(\ref{eq:S_lsf}) at 1000 K and adding the corresponding entropy
term to the total energy of the alloy. 

Although the LSF energy of Ni does not have the exact quadratic
form, Eq. (\ref{eq:S_lsf}) still provides reasonable
description of the LSF entropy. For instance, it yields 0.70
$\mu_B$ for the local magnetic moment of Ni
in Fe$_{65}$Ni$_{35}$  at 1000 K, while a more accurate result
from (\ref{eq:m_lsf}) is 0.82 $\mu_B$. At the same time,
it is clear that (\ref{eq:S_lsf}) somewhat underestimates
the magnetic entropy related to the LSF on Ni atoms in 
Fe$_{65}$Ni$_{35}$ and consequently the magnetic moment
of Ni in the paramagnetic state at high temperatures.

The LSF energy of Fe is quite different from that of Ni:
its minimum is at 1.77 $\mu_B$, although it is quite shallow
with a steep increase beyond 2.5 $\mu_B$. Although the LSF
energy of Fe does not resemble parabola, the magnetic
moment of Fe due to the LSF is about 1.8 $\mu_B$ at 1000 K,
independently if (\ref{eq:S_lsf}) or (\ref{eq:m_lsf}) are
used in the calculations. One can also see, that the 
LSF affect little the magnitude of magnetic moment.
However, as will be shown below, the magnetic entropy of Fe
affects substantially the thermal lattice
expansion in the paramagnetic state.

\subsection{Magnetic exchange interactions in Fe$_{65}$Ni$_{35}$}

Magnetic state given by the corresponding spin-spin correlation
functions, and its temperature dependence are key parameters
needed for a quantitatively accurate modeling of the Invar effect.
Although there exist experimental data on the reduced
magnetization,\cite{crange63} nothing is
known about MSRO or spin-spin correlation functions in Fe-Ni
alloys. Therefore the only way to get this information is
to use theoretical simulations.

We assume that the magnetic energy and spin configuration
of the Fe-Ni Invar alloys at particular temperature can be determined
in statistical thermodynamics simulations using the following
classical Heisenberg Hamiltonian:

\begin{equation} \label{eq:H_H_all}
H = - \sum_p \sum_{i,j \in p } \sum_{\alpha, \beta = \rm Fe,Ni}
J_p^{\alpha \beta} c_i^{\alpha}c_j^{\beta} \mathbf{e}_{i} \mathbf{e}_{j} .
\end{equation}
Here, $J_p^{\alpha \beta}$ are the magnetic exchange interactions
between $\alpha$ and $\beta$ alloy components for
coordination shell $p$ and $\mathbf{e}_{i}$ is the 
direction of the spin at site $i$;
$c_i^{\alpha}$ takes on value 1 if site $i$ is occupied by 
atom $\alpha$ and 0 otherwise.

Of course, the Fe-Ni Invar alloys are not a Heisenberg system:
the corresponding magnetic exchange interactions depend not
only on the local and global magnetic state but also on the
local chemical environment.\cite{ruban05} However, the dependence
on the local environment is strongest in the completely ordered FM
state and can be neglected in  qualitative statistical 
thermodynamics simulations at elevated temperatures. 
Therefore we calculate magnetic exchange interactions in
random alloys within the CPA using magnetic force theorem\cite{liecht87}
as is implemented in the Lyngby version of the EMTO code\cite{ruban16}
within the LDA.\cite{j_xc_LDA} 

\begin{figure}[tbp]
\includegraphics[height=10.0cm]{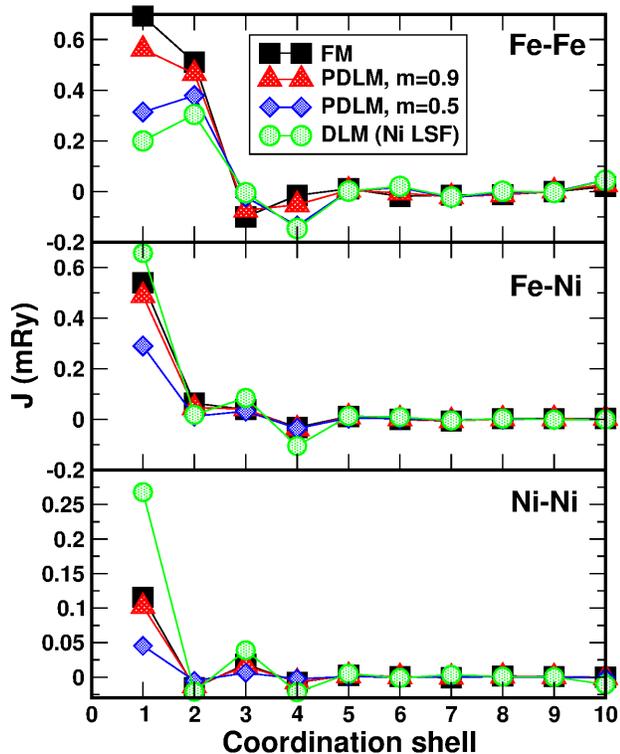}
\caption[5]{(Color online) Magnetic exchange interactions in
random Fe$_{65}$Ni$_{35}$ alloy in the FM state with different
magnetizations and in the paramagnetic state with LSF on Ni
at 500 K, which is the Curie temperature of Fe$_{65}$Ni$_{35}$. 
}
\label{fig:J_xc}
\end{figure}

As has been already mentioned, the magnetic exchange interactions
in Fe-Ni alloys depend quite strongly on the magnetic state,
as is also the case of magnetic exchange interactions in bcc Fe.\cite{ruban04} 
In Fig. \ref{fig:J_xc}, we show the magnetic exchange interactions
in Fe$_{65}$Ni$_{35}$ random alloy as a function of magnetization,
which have been determined in the CPA-PDLM model calculations
in the same way as in Ref. \onlinecite{ruban08}. In this case, the
first-principles CPA calculations have been done for the
following model alloy configuration:
(Fe$\uparrow_{y}$Fe$\downarrow_{1-y}$)$_c$Ni$_{1-c}$, 
where $y$ is connected to the reduced magnetization, $m$, as
$m = 1 - 2y$, and $c$ is the concentration of Fe.\cite{ruban08}
That is, only Fe atoms are used for the modeling of the reduced
magnetization, while Ni atoms are left to acquire the magnetic
moment according to the self-consistent calculations.

One can see that Fe-Ni and Ni-Ni magnetic exchange interactions
decreasing together with decreasing magnetization, and in
the paramagnetic state, when $m=0$, they should vanish in this
model, since Ni magnetic moment disappears in the DLM-CPA
calculations.
This is not, however, the case if one consider the LSF on
Ni at finite temperature. In Fig. \ref{fig:J_xc},
we show the Fe-Ni and Ni-Ni magnetic exchange interactions
at 500 K (which is the Curie temperature of Fe$_{65}$Ni$_{35}$)
due to the LSF on Ni. The latter were taking into consideration
using (\ref{eq:S_lsf}) for the LSF entropy on Ni during DFT 
self-consistent calculations. As a result, the Fe-Ni and 
especially Ni-Ni magnetic exchange interactions at the first
coordination shell become substantially larger than those
in the FM state. This produces a pronounced effect on the
Cuire temperature as will be demonstrated below.

\subsection{Magnetic transition in Fe-Ni alloys}
\label{sec:HMC}

The dependence of magnetic exchange interactions on the
global magnetic state can be neglected in the calculations
of the magnetic phase transition, which being the second
order happens practically in the paramagnetic state
(although with a large amount of the MSRO).
In this case, one can chose only one set of magnetic
interactions, which corresponds to the paramagnetic state
just above the Curie temperature, and do statistical
thermodynamics simulations of the magnetic transition.
This is the corresponding state approach,\cite{ruban04} which
has been, for instance, used before in the Curie temperature
calculations of bcc Fe. 

Here, we determine magnetic exchange interactions for the
lattice constant and thermal one-electron excitations, which
correspond to the experimental Curie temperature.\cite{crange63,hayase73}
The use of the experimental Curie temperature in the 
calculations of the magnetic exchange interactions 
also simplifies the account for the LSF.
As has been demonstrated above, the LSF on Ni lead to a
substantial renormalization of the magnetic exchange interactions.

We have used two different schemes to account for the
LSF on Ni: 1) DFT self-consistent calculations
using (\ref{eq:S_lsf}) for the LSF entropy on Ni at a given
temperature and
2) the use of the single-site mean-field approximation
(\ref{eq:m_lsf}) for the average magnetic moment of Ni
at a given temperature from the corresponding LSF energy.
Although the second approach is quite time consuming, it is
more accurate and this is important since magnetic exchange
interactions are roughly proportional to the magnitude of the
local magnetic moment.

In order to demonstrate that the described above method
works reasonably well, we calculate the magnetic phase transition
in several Fe-Ni alloys covering the whole concentration range of 
the fcc Fe-Ni alloys, including pure Ni.  All the calculations have
been done for random alloys using the CPA. The latter is probably a
rough approximation for Ni-rich Fe-Ni alloys, where the Curie
temperature is quite close to the atomic order-disorder phase
transition. For instance, the Curie temperature of Fe$_{25}$Ni$_{75}$
is just about 100 K above the order-disorder phase transition.
Nevertheless, we disregard the atomic short-range ordering 
going mainly after a qualitative picture. 

\begin{table*}
\caption{Calculated local magnetic moments on Ni in
paramagnetic state due to LSF at the experimental Curie temperature
and lattice constants\cite{crange63,hayase73} from (\ref{eq:S_lsf})
and from (\ref{eq:m_lsf}) and theoretical and experimental
Curie temperatures of some Fe-Ni alloys including pure Ni.
The results obtained in the FM state are also shown for
comparison.}
\begin{ruledtabular}
\begin{tabular}{rrrrr}
Alloy  & Fe$_{65}$Ni$_{35}$ & Fe$_{50}$Ni$_{50}$ & Fe$_{30}$Ni$_{70}$ & Ni \\  
\hline
 m$_{\rm Ni}$ FM                              & 0.75  & 0.71  & 0.66   & 0.62 \\  
 m$_{\rm Ni}$ DLM-LSF;(\ref{eq:S_lsf}) for Ni & 0.52  & 0.65  & 0.70   & 0.68 \\
 m$_{\rm Ni}$ DLM-LSF;(\ref{eq:m_lsf}) for Ni & 0.66  & 0.69  & 0.66   & 0.58 \\ 
\hline
T$_c$: J$_{xc}$ FM                              &  520  &   560   &	 540 &  310  \\
T$_c$: J$_{xc}$ DLM-LSF;(\ref{eq:S_lsf}) for Ni &  360  &   630   &	 810 &  810  \\
T$_c$: J$_{xc}$ DLM-LSF;(\ref{eq:m_lsf}) for Ni &  430  &   650   &	 800 &  710  \\
\hline
T$_c$ Experiment	                            &  500  &   780   &	 880 &  630  \\
\end{tabular}
\end{ruledtabular}
\label{tbl:T_c}
\end{table*}

The calculated and experimental Curie temperatures are presented
in Table \ref{tbl:T_c}. The theoretical Curie temperatures
have been obtained in the Heisenberg Monte Carlo simulations
using a simulation box of 12$\times$12$\times$12($\times$4)
on the fcc underlying lattice. The transition temperature was
determined approximately from the maximum of the heat capacity.
Since the temperature step was 10 K, this is a kind
of an error bar for the theoretical Curie temperatures.

One can see, that the Curie temperature is indeed sensitive to the
magnitude of the Ni local magnetic moment. The overall best agreement
of the calculated Curie temperature with experimental data is obtained
using the magnetic moment from single-site mean-field
modeling (\ref{eq:m_lsf}).
As one can see in Table \ref{tbl:T_c}, the LSF entropy given by
Eq. (\ref{eq:S_lsf}) underestimates the induced magnetic moment of
Ni in the Fe-rich alloys and overestimates it in the Ni-rich alloys
and pure Ni. 

It is interesting, that the magnetic exchange interactions
in the completely ordered FM state yield the best results
for the Curie temperature of the Invar Fe$_{65}$Ni$_{35}$
alloy (see Table \ref{tbl:T_c}). However, the FM interactions
do not reproduce the general trend of the concentration
dependence of the Curie temperature in Fe-Ni alloys with
maximum around 70 at.\% of Ni. This is so since these
alloys are not Heisenberg systems, and thus the FM exchange
interactions are hardly relevant to the magnetic exchange
interactions at the Curie temperature close to the
paramagnetic state.

\section{Spin-wave method results for the 0 K total energy}

\subsection{Completely ordered ferromagnetic and ideal paramagnetic
states}

To get a reasonably accurate account for the MSRO, the integration
over q-points in the SWM (see Eq. (\ref{eq:E_fm})) was done using
the 7$\times$7$\times$7 Monkhorst-Pack grid,\cite{monkhorst72} which
results in 20 non-equivalent q-points in the irreducible part of
the fcc Brillouin zone. For every Wigner-Seitz (WS) radius within
the range of 2.6 -- 2.7 a.u. (with the step of 0.005 a.u.), the total
energies of Fe$_{65}$Ni$_{35}$ alloy were calculated in the
corresponding 20 PSS states by the EMTO-CPA method (see Appendix
\ref{NC_SS} and \ref{SWM} for details).  
Then, for every WS radius, the total energy of Fe$_{65}$Ni$_{35}$
in the given magnetic state was obtained by the corresponding weighting
of the total energies obtained in the PSS calculations (see
Appendix \ref{SWM}).

The total energy of the completely ordered ferromagnetic state is,
of course, just the total energy of the PSS with $\mathbf{q} = 0$.
The total energy of the IPM state is given by the equal weighted 
sum of the total energies of all spin-spirals. 
Another way to calculate the total energy of the IPM state is to do
collinear DLM-CPA calculations.
For a Heisenberg system these two methods should produce
the same energy.

\begin{figure}[tbp]
\includegraphics[height=8.0cm]{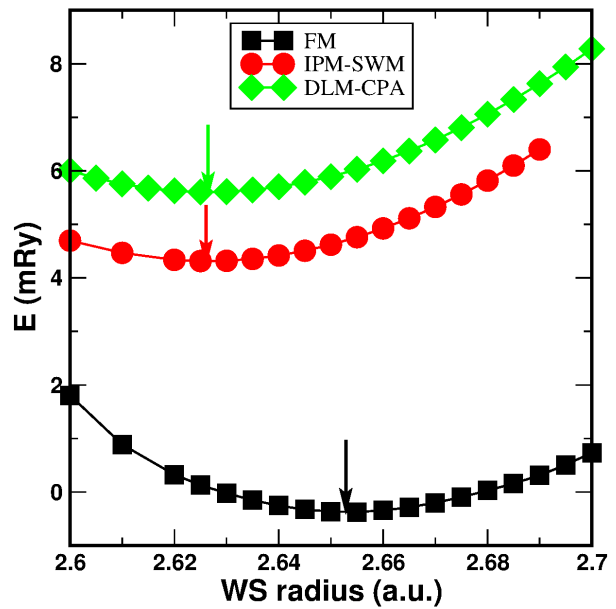}
\caption[1]{(Color online) Total energy of random Fe$_{65}$Ni$_{35}$
in the FM and PM states. The latter is obtained in the CPA-DLM (diamonds)
and SWM (circles) calculations. The arrows show the position of
equilibrium WS radii.}
\label{fig:E_tot}
\end{figure}

However, it is not the case of Fe$_{65}$Ni$_{35}$. As is seen 
in Fig. \ref{fig:E_tot}, the energy of the IPM state in the SWM is 
1 mRy lower than the DLM-CPA energy. At the same time, the FM
energy is only about 5 mRy below the energy of the paramagnetic
state. The reason of such disagreement between the DLM-CPA and SWM
total energies is a pronounced itinerant electron character of magnetism
in this alloy.
It is also reflected in the difference of the local magnetic
moments of Fe and Ni in the IPM state. While the local magnetic moment
of Ni vanishes in the DLM-CPA calculations, its average magnitude in
the SWM is about 0.1 $\mu_B$.

\begin{figure}[tbp]
\includegraphics[height=8.0cm]{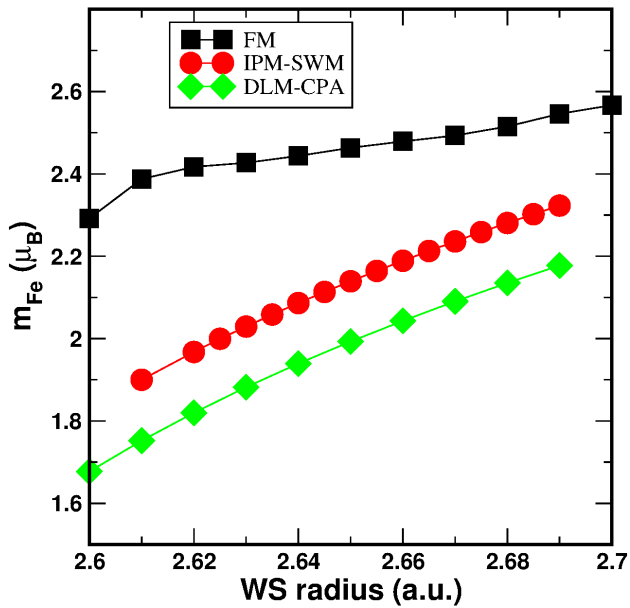}
\caption[1]{(Color online) Local magnetic moment of Fe in
Fe$_{65}$Ni$_{35}$  as a function of the WS radius in the FM and
paramagnetic states.}
\label{fig:m_Fe}
\end{figure}

The latter is due to the contributions from long wave PSS.
The magnetic moment of Fe in the SWM calculations is also larger than
that obtained in the DLM-CPA calculations for the same reason.
At the same time, the magnetic moment of Fe in the paramagnetic
state is substantially smaller than one in FM state, as one
can see in Fig. \ref{fig:m_Fe}. 

Let us note that in spite of all the above mentioned differences in
results for energies and magnetic moments, the SWM and DLM-CPA agree
on the equilibrium
volume or WS radius, of the IPM state. The later is about 2.626 a.u.
in both cases (which corresponds to the lattice constant of 3.554 \AA).
The equilibrium WS radius in the FM state, $S^{\rm FM}$ is 2.653 a.u.
(lattice constant is 3.593 \AA, without contribution from zero
point lattice vibrations). This is exactly what was reported in the
previous EMTO calculations.\cite{ruban07}

\subsection{Magnetic short-range order effects}

\begin{figure}[tbp]
\includegraphics[height=8.0cm]{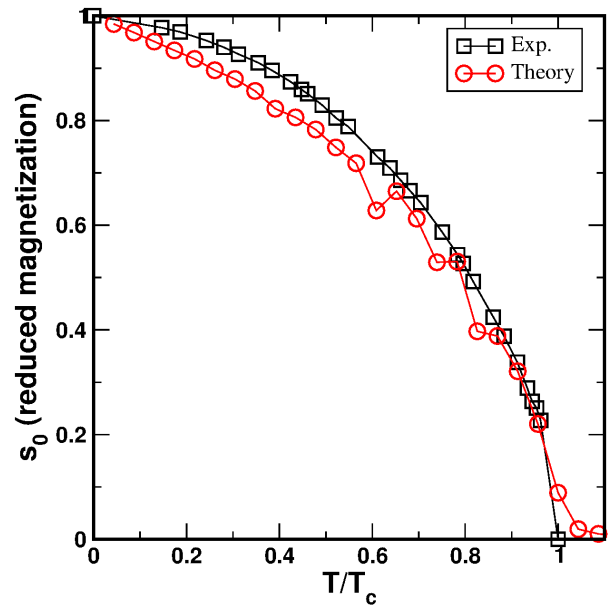}
\caption[6]{(Color online) Reduced magnetization as a function of
reduced temperature in Fe$_{65}$Ni$_{35}$ : Experiment\cite{crange63}
vs present theoretical modeling.}
\label{fig:m_T}
\end{figure}

In the SWM, the MSRO effects are taken into consideration through
the corresponding spin-spin correlation functions. In this work, 
the spin-spin correlation functions have been obtained in the Heisenberg
Monte Carlo simulations described in Sec. \ref{sec:HMC}.
Although those calculations have been done for a fixed set
magnetic exchange interactions which correspond to the
paramagnetic state, the calculated reduced magnetization of
Fe$_{65}$Ni$_{35}$ turns out to be in good agreement with
the experimental one \cite{crange63} in the whole temperature
range, as is shown in Fig. \ref{fig:m_T}. Therefore we expect
that the theoretical results for the spin-spin correlation functions
are qualitatively correct and can be used in the SWM modeling.

Since the homogenous magnetic state is assumed in the single-site CPA
spin-spiral calculations of Fe$_{65}$Ni$_{35}$, the component-resolved
spin-spin correlation functions from the Monte Carlo simulations should
be reduced to the average ones consistent with the first-principles
spin-spiral calculations in order to be used in the SWM.
For the completely random alloy, the average spin-spin correlation
function is

\begin{eqnarray}Ê\label{eq:xi_av}
\tilde{\xi}(\mathbf{R}) &=& c^2 \xi^{\rm Fe-Fe}(\mathbf{R}) +
2 c(1-c) \xi^{\rm Fe-Ni}(\mathbf{R}) \\ \nonumber
&+& (1-c)^2 \xi^{\rm Ni-Ni}(\mathbf{R}) ,
\end{eqnarray}
where $\xi^{\rm Fe-Fe}(\mathbf{R})$, $\xi^{\rm Fe-Ni}(\mathbf{R})$,
and $\xi^{\rm Ni-Ni}(\mathbf{R})$  are the Monte Carlo
results for the spin-spin correlation functions of different 
alloy pairs.

In fact, the difference between these three alloy-component resolved 
spin-spin correlation functions is quite small in the FM state, and
becomes pronounced only at high temperatures, especially for distant
coordination shells. However, in the latter case, the correlation functions
themselves become quite small. For instance, the largest nearest neighbor
spin-spin correlation functions at 300 K are 0.820, 0.750, and 0.677,
while at 1000 K they are 0.091, 0.070, and 0.024 for Fe-Fe,
Fe-Ni, and Ni-Ni pairs, respectively. This means that averaging
(\ref{eq:xi_av})  does not introduce a noticeable error.

\begin{figure}[tbp]
\includegraphics[height=8.0cm]{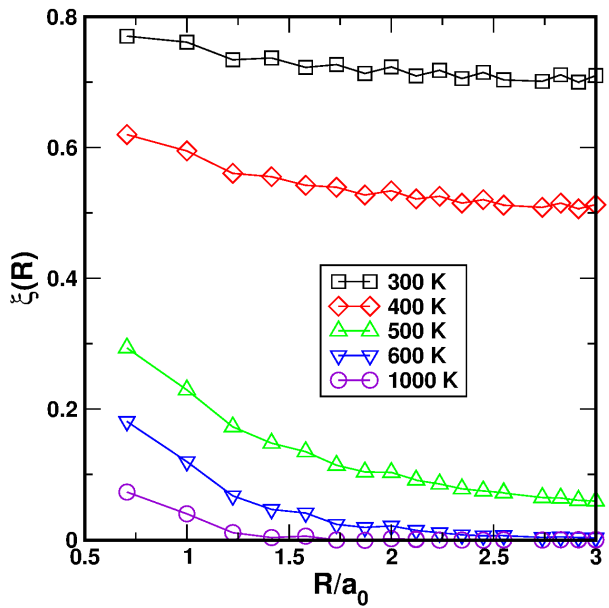}
\caption[7]{(Color online) Average spin-spin correlation function
in Fe$_{65}$Ni$_{35}$ at different temperatures.}
\label{fig:xi_R}
\end{figure}

The spatial behavior of the spin-spin correlation functions is
shown in Fig. \ref{fig:xi_R}. In the totally ordered FM state,
$\tilde{\xi}(\mathbf{R})=1$, and in the IPM ,$\tilde{\xi}(\mathbf{R})=0$.
In the PDLM model for the FM state with reduced magnetization
$s_0$, $\tilde{\xi}(\mathbf{R})=s_0$. One can see, that the
MSRO becomes important close to the magnetic phase transition,
exactly where the Invar effect is observed.

The average spin-spin correlation functions have been used
in the SWM calculations in order to get the total energy of 
Fe$_{65}$Ni$_{35}$ at a given temperature (see Appendix \ref{SWM}
for details). Since the
calculated Curie temperature and theoretical one differs,
the theoretical temperature dependence of the spin-spin
correlation functions has been rescaled in order to have the Curie
temperature and MSRO consistent with the experiment.\cite{crange63} 

\begin{figure}[tbp]
\includegraphics[height=8.0cm]{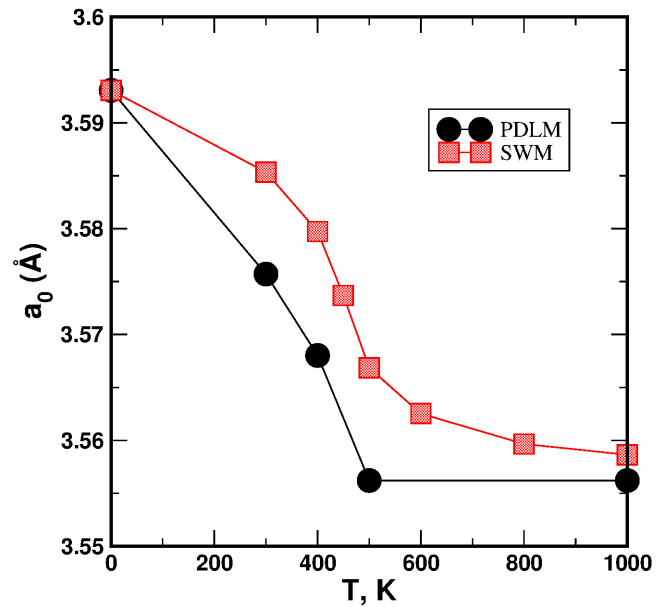}
\caption[8]{(Color online) Zero temperature lattice constant
obtained in the SWM and PDLM calculations as a function of magnetic
state translated to the corresponding temperature shown on
the x-axis.}
\label{fig:a_0}
\end{figure}

In Fig. \ref{fig:a_0}, we show the results for the 0 K equilibrium
lattice constant of Fe$_{65}$Ni$_{35}$ obtained in the SWM calculations
from the spin-spin correlation functions at the corresponding temperature,
which is shown on the x-axis. On the same figure, we also show the
results of the PDLM-CPA calculations, where the mapping to the
temperature is done only according to the reduced magnetization
since the MSRO effects are absent in the PDLM-CPA model. The latter
is, in particular, the reason for the abrupt change of the 
lattice constant at 500 K, which is the Curie temperature,
and its constant value above the magnetic phase transition.
It is clear that the  MSRO effects are pronounced and
cannot be neglected in the accurate \textit{ab initio} modeling of
the Invar effect.

\section{Calculated thermal lattice expansion} 

\subsection{Debye-Gr\"uneisen model}

As has been mentioned above, the Debye-Gr\"uneisen
model\cite{moruzzi88,korzhavyi94} is the
only way to account for the thermal lattice expansion in
the FeNi Invar alloys at present time. A combination
of chemical randomness with highly non-trivial thermal electronic
and magnetic excitations leave no chance, for instance,
for using quasiharmonic approximation based on the \textit{ab initio}
phonon calculations.

Three parameters are needed for the Debye-Gr\"uneisen modeling: bulk
modulus, Gr\"uneisen constant, and zero-temperature equilibrium
lattice constant. However, it is quite difficult to obtain stable
results for the  Gr\"uneisen constant, especially in the FM state,
where the magnetic moment of Fe does not change linearly
(see Fig. \ref{fig:m_Fe}).

In the present study, the Morse fit\cite{moruzzi88} was
used for a parametrization of the total energy (or the Helmholtz
free energy) as well as for determining the parameters of
the Debye-Gr\"uneisen model, which seemed to be quite stable and
not so sensitive to the range of energy fitting, the size of the
step, and other details. But even in this case, the calculated 
Gr\"uneisen constant exhibited some fluctuations related to
the details of calculations. So, in the end, in order to simplify
the tedious numerical exercise, its value was fixed to 1.8,
which is close to the obtained values also by Liot
\textit{et al.}\cite{liot09}, for
all the Debye-Gr\"uneisen calculations, which were done
using the formalism outlined in Ref. \onlinecite{korzhavyi94}.

\subsection{Transverse and longitudinal spin fluctuations in the 
ideal paramagnetic state}

\begin{figure}[tbp]
\includegraphics[height=8.0cm]{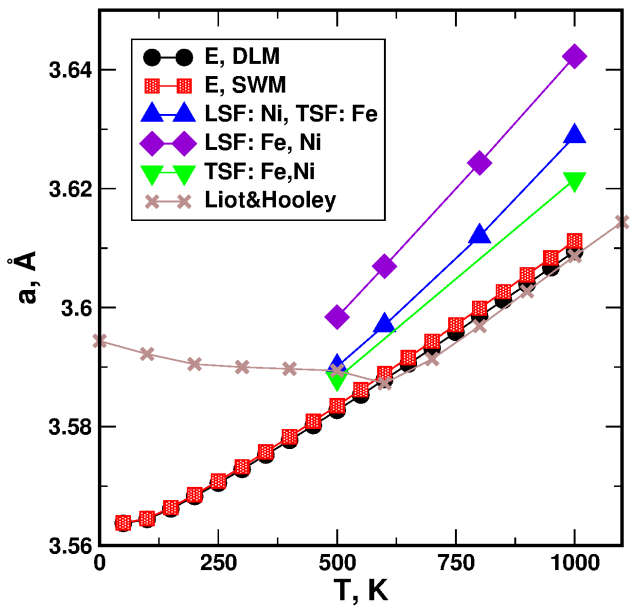}
\caption[9]{(Color online) Temperature dependence of the
lattice constant of the IPM of Fe$_{65}$Ni$_{35}$ obtained in the
Debye-Gr\"uneisen model using (1) the total energy of from the DLM-CPA
calculations (E, DLM); (2) the total energy from the SWM calculations
(E, SPW); The other results are obtained using the Helmholtz free energy
at the corresponding temperatures with contributions from the one-electron
excitations, LSF using (\ref{eq:S_lsf}) and TSF using (\ref{eq:S_TSF})
in different combinations. Liot and Hooley EMTO-CPA results from Ref. \onlinecite{liot12}.
}
\label{fig:a_dlm_T}
\end{figure}

Before we discuss the effect of the MSRO, we would like to demonstrate
the effect of spin fluctuations on the temperature dependence of the lattice
constant of Fe$_{65}$Ni$_{35}$ in the IPM. The starting point here is
the total energy of the IPM state without contributions from the electronic
and magnetic thermal excitations. In Fig. \ref{fig:a_dlm_T}, the
corresponding lattice constants obtained in the Debye-Gr\"uneisen
model from the DLM-CPA and SWM total energies (E) are shown. They
agree well with each other and with the results by Liot and
Hooley\cite{liot12} obtained in the similar DLM-CPA calculations.

The other results shown in Fig. \ref{fig:a_dlm_T} are obtained in
the Debye-Gr\"uneisen model using the Helmholtz free energy, which
includes contributions from the one-electron and magnetic excitations
at the corresponding temperature. 
Three different combinations have been considered and shown in
Fig. \ref{fig:a_dlm_T}: 1) LSF on Ni and TSF on Fe using entropies
(\ref{eq:S_lsf}) and (\ref{eq:S_TSF}), respectively (LSF: Ni, TSF: Fe);
2) LSF on both Fe and Ni (LSF: Fe,Ni); and
3) TSF on both alloy components, although in effect, it is only on
Fe, since the local magnetic moment on Ni vanishes in this case
(TSF: Fe,Ni).

As one can see, the entropy contribution produces a considerable effect
on the lattice constant. Even the inclusion of only TSF on Fe (TSF:Fe,Ni)
leads to a noticeable change of the lattice constant at high temperature.
The reason, why the TSF on Fe play so important role in the thermal
expansion of Fe$_{65}$Ni$_{35}$ in the paramagnetic state, can be
traced back to the quite steep increase of the local magnetic moment
of Fe with the lattice constant as is seen in Fig. \ref{fig:m_Fe}.

The addition of the LSF on Ni (LSF: Ni, TSF: Fe) also increases
the lattice constant at high temperatures, but to a less degree
than TSF on Fe. Even more drastic increase of the high temperature
lattice constant is obtained with LSF on both Fe and Ni (LSF: Fe,Ni).
As has been mentioned above, there is no doubt that LSF are very much
relevant for Fe too, since its magnetic moment is quite sensitive
to the lattice constant and magnetic state.

However, the LSF energy of Fe, which has minimum at 1.77
$\mu_B$, does not resemble parabola and therefore
the use of (\ref{eq:S_lsf}) here should be considered as a very
rough estimate. Nevertheless, it is clear, that 
the LSF on Fe should be considered too in the accurate modeling of
the Invar effect in Fe$_{65}$Ni$_{35}$. This point will be also
clear in the next section, where the effect of the magnetic
long- and short-range order is considered.

\subsection{Contribution from magnetic long- and short-range order
effects}

\begin{figure}[tbp]
\includegraphics[height=8.0cm]{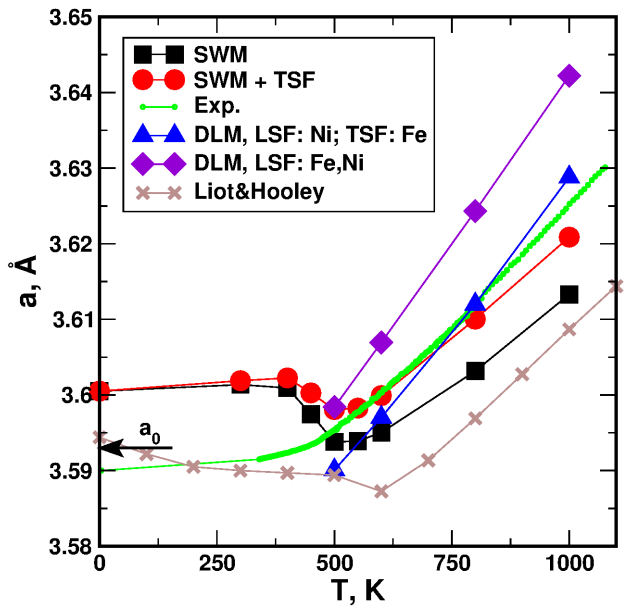}
\caption[10]{(Color online) Lattice constant of Fe$_{65}$Ni$_{35}$
obtained from the SWM total energy calculations (SWM) without
and with the PDLM entropy (SWM + S$^{\rm PDLM}$), from DLM
calculations with LSF on Ni and TSF on Fe and LSF on
both elements. The experimental data are taken from Ref.
\onlinecite{gorria09}. Liot and Hooley
EMTO-CPA results from Ref. \onlinecite{liot12}.
The arrow ($a_0$) indicates
the equilibrium lattice constant obtained in the FM state
without zero point lattice vibration contribution.
}
\label{fig:a_T}
\end{figure}

As has been demonstrated above (see Fig. \ref{fig:a_0}), the
inclusion of the MSRO leads to a noticeable increase of the 0 K 
lattice constant, especially in the case of magnetic states in
the vicinity of the Curie temperature.
Consequently, this results in the corresponding shift of the 
equilibrium lattice constants at finite temperatures obtained in
the Debye-Gr\"uneisen model. In Fig. \ref{fig:a_T}, we show
the SWM results, which include the MSRO effects,
and the PDLM-CPA results by Liot and Hooley\cite{liot12} where
the effect of the MSRO is absent. Clearly, the MSRO leads to 
a substantial shift of the lattice constant. The inclusion of the
MSRO effects changes also the slope of the temperature dependence
of the lattice constant in the temperature range between 0 and
400 K from negative in the PDLM-CPA calculations to positive, 
thereby making theory consistent with the experimental data.

As has been demonstrated in the previous section, one has also
to account for the magnetic entropy contribution in quantitative
analysis of the thermal lattice expansion in Fe-Ni alloys.
Fortunately, it can be easily done within the SWM for the TSF since
the magnitude of the magnetic moment entering (\ref{eq:S_TSF}) is just
the average magnitude of the magnetic moment in the SWM calculations.
Then the TSF magnetic entropy in the SWM in the presence of
magnetic long- and short-range order can be determined using
a PDLM-like single-site mean field approximation, which is
consistent with the corresponding definition of the total
energy in the SWM (see eq. (\ref{eq:E_fm})):

\begin{equation} \label{eq:S_PDLM}
S^{\rm PDLM} = (1-s_0) \sum_i c_i\ln(1 + m_i) .
\end{equation}
Here, $s_0$ is the reduced magnetization, $c_i$ and $m_i$ 
are the concentration and local average magnetic moment of alloy
component $i$. In Heisenberg systems with localized type of magnetism,
the local magnetic moments are constant, but it is not the case
of Fe-Ni alloys, and therefore they have been determined as the
average over all the magnetic moments of spin-spirals except
for $\mathbf{q}=0$ weighted according to the given MSRO.

Equation (\ref{eq:S_PDLM}) does not account, however, for the MSRO.
In an approximate way, it can be done by using $s_0 = \xi_1$, i.e.
the value of the spin-spin correlation function at the first
coordination shell. This, at least, provides a smooth behavior
of the entropy as a function of magnetic state near the Curie
temperature. 

The lattice constant of Fe$_{65}$Ni$_{35}$ with MSRO and TSF
is shown in Fig. \ref{fig:a_T} (SWM + TSF). Although the account of
the TSF improves the results for the lattice constant near the Curie
temperature (500 K), still the lattice constant has a pronounced minimum
in this region. The experimental lattice constant, in contrast, exhibits
a very smooth temperature dependence with two different slopes below and
above the Curie temperature. It is clear, that something is still
missing in theory and this is most probably the LSF, the proper account
of which can lift up the lattice constants in the paramagnetic 
region and eliminate the minimum at the Curie temperature.

The point is that the SWM results are in fact in good agreement with
experimental data up to 400 K in the FM region, but shifted up by
approximately 0.01 \AA.  In the paramagnetic region, this shift
however disappears and the inclusion of the TSF helps but little.
An additional shift can be obviously obtained from the LSF, which
is obvious if one compares the DLM+LSF results with the DLM results
by Liot and Hooley in Fig. \ref{fig:a_T}.
The LSF are small in the FM state and practically disappear 
with increasing magnetization. And vice versa, with decreasing
magnetization, while approaching the magnetic transition, the role
of the LSF increases. Therefore one can speculate that they are
important in order to reproduce the smooth temperature dependence
of the lattice constant in Fe$_{65}$Ni$_{35}$ near the Curie
temperature.

Thus, both contributions, MSRO and LSF, are important for a quantitatively
correct description of the thermal expansion in Fe-Ni alloys.
The problem is, however, that it is not clear how to combine the LSF
formalism with the SWM, which seems to be at the moment the only
practical theoretical tool, which allows the inclusion of the MSRO
effects in random alloys. Most probably, the solution requires
the development of other types of techniques, which effectively 
combine these two finite temperature features of the itinerant
magnets.

\section{Summary}

1. The spin-wave method is generalized here to the
ferromagnetic state. 

2. The developed SWM in combination with the CPA spin-spiral
calculations is used then to calculate the contribution from the
magnetic long- and short-range order to the total energy of the
Fe$_{65}$Ni$_{35}$ Invar alloy. 

3. The calculations show that the MSRO effects yield significant
contribution to the temperature dependence of the lattice constant.
In particular, they correct the PDLM-CPA results in the FM region
up to 400 K, producing the thermal lattice expansion consistent
with the experimental data. 

4. The entropy contribution associated with the LSF and TSF has
been considered in the ab initio modeling of the Invar effect.
It provides quite significant shift of the lattice constant at
high temperatures in the paramagnetic state, and should be
considered in the accurate modeling of the thermal expansion
in Fe-Ni alloys.

5. Although both the MSRO effect and LSF are important for
accurate calculations of the equilibrium lattice constant of
Fe-Ni alloys at elevated temperatures, it is practically impossible
to include them together in the modeling within the used here
theoretical techniques. This results in an abnormal behavior of
the temperature dependence of the lattice constant close to the
Curie temperature.

6. The magnetic transition temperature in the fcc Fe-Ni alloys is
calculated taking the LSF on Ni into consideration. It is demonstrated
that the LSF on Ni provide a substantial strengthening of
the Ni-Ni and Fe-Ni magnetic exchange interactions in the
paramagnetic state.

%%%%%%%%%%%%%%%%%%%%%%%%%%%%%%%%%%%%%%%%%%%%%%%%%%%%%%%%%%%%%%%%%%%%%%%

\appendix
\section{Spin-spiral and noncollinear calculations by the EMTO-CPA
method}
\label{NC_SS}

The present implementation of noncollinear and spin-spiral calculations
in the EMTO-CPA Green's function method is done within the atomic
sphere approximation where the magnetic configuration is site-projected,
with the magnetic moment of a site given by the integrated magnetic
density within the corresponding atomic sphere:

\begin{equation} \label{m(R)}
\mathbf{m}_i = \int_{\Omega_i} d\mathbf{r} \:
\mathbf{m}(\mathbf{r} - \mathbf{R}_i) .
\end{equation}
Here, $\mathbf{m} ( \mathbf{r} - \mathbf{R}_i )$ is the magnetic density
centred at site $i$ in position $\mathbf{R}_i$, and $\mathbf{m}_i$
is the magnetic moment of site $i$. 

The magnetic configuration of the system is then given by polar
angels $\theta_i$ and $\phi_i$ of the local magnetic moments
$\mathbf{m}_i = m_i \,
( \sin\theta_i \cos\phi_i, \, \sin\theta_i \sin\phi_i, \, \cos\theta_i )$
of each site relative to the $z$-axis of the global frame of reference. 
If the local magnetic moment $m_i$ is defined in the local 
frame of reference along the local $z$ direction, it transforms to the
global frame by the corresponding spin-$\frac{1}{2}$ rotation
matrix:\cite{sandratskii98,mryasov91}

\begin{eqnarray} \label{eq:U}
\mathbf{U}_i = \begin{bmatrix}
    \cos(\theta_i/2) e^{i\phi_i/2} & \sin(\theta_i/2) e^{-i\phi_i/2} \\
    \\
   -\sin(\theta_i/2) e^{i\phi_i/2} & \cos(\theta_i/2) e^{-i\phi_i/2} \\
      \end{bmatrix}
\end{eqnarray}

As has been demonstrated by Mryasov {\it et al.} \cite{mryasov91},
in the site-projected formalism, the electronic structure of a
non-collinear system in the KKR or LMTO methods can be obtained
using the local frame references of individual sites by the following
transformation of the structure constants, or slope
matrix\cite{andersen94,andersen98} in the case of the EMTO method:

\begin{equation}
\mathbf{\tilde{S}}^{ij}_{LL'} =
\mathbf{U}^{-1}_i \mathbf{S}^{ij}_{LL'} \mathbf{U}_j .
\end{equation}
Here, $\mathbf{S}^{ij}_{LL'} = \mathbf{I} S^{ij}_{LL'}$
and $S^{ij}_{LL'}$ is the slope or structure constants matrix of
the EMTO method and $\mathbf{I}$ is the unit matrix.
Since $\mathbf{U}_i$ is unitary,
$\mathbf{U}^{-1}_i = \mathbf{U}^{\dagger}_i$.

A similar formalism is valid in the case of spin-spiral
calculations.\cite{mryasov91} Here, it is outlined for 
Bravais lattices in order to simplify notations, since
its generalization to the multisite systems is straightforward.
The spin spiral magnetic configuration for a fixed azimuth angle
$\theta$ and wave vector $\mathbf{q}$ is determined by

\begin{eqnarray}
\mathbf{m}_i = m_i \, \left( \sin\theta\cos(\mathbf{q}\mathbf{R}_i),\sin\theta\sin(\mathbf{q}\mathbf{R}_i),\cos\theta \right) ,
\end{eqnarray}
and the corresponding site projected spin rotation matrix
(\ref{eq:U}) is obtained by substitution $\phi_i = \mathbf{q}\mathbf{R}_i$.

In the absence of the spin-orbit interaction, one can use
the generalized Bloch theorem,\cite{sandratskii98,mryasov91} which
allows one to disentangle the magnetic and crystal structures translational
symmetries thereby reducing electronic structure calculations
to the primitive unit cell of the underlying magnetic structure
Bravais lattice. A spin spiral magnetic structure in this case is
set up by the
corresponding transformation of the structure constants (or slope
matrixes). In particular, the structure constants or slope matrix
for a spin-spiral with wave vector $\mathbf{q}$ and azimuth angle
$\theta$ is

\begin{eqnarray} \label{eq:S_ss}
&&\mathbf{S}_{LL'}(\mathbf{k}) = \\ \nonumber
&& \mathbf{U}^{-1}\begin{bmatrix}
    S_{LL'}(\mathbf{k} - \frac{\mathbf{q}}{2}) & 0 \\
    \\
    0 & S_{LL'}(\mathbf{k} + \frac{\mathbf{q}}{2})\\ 
      \end{bmatrix} \mathbf{U} ,
\end{eqnarray}
where spin transformation matrix $U$ is given by (\ref{eq:U})
with $\phi = 0$. 

The potential function of the EMTO method,
$\mathbf{D}_{L}(z)$,\cite{andersen94,andersen98} for particular
energy $z$ and angular magnetic moments $L=(l,m)$ 
is obtained in the local frame of reference and thus it is a diagonal
matrix in the spinor basis with "spin-up" and "spin-down"
components: $D^{(\frac{1}{2})}_{L}(z)$ and 
$D^{(-\frac{1}{2})}_{L}(z)$. The path operator, whose poles are
the electronic spectrum of the system and which is the main
quantity for calculating electron density in the Green's function
formalism is then

\begin{eqnarray} \label{eq:path_op}
\mathbf{g}_{LL'}(\mathbf{k},z) =
\frac{1}{\mathbf{S}_{LL'}(\mathbf{k},z) - \mathbf{D}_{L}(z)} .
\end{eqnarray}

The above formalism is in fact straightforwardly generalized to
the case of random alloys within the single-site coherent potential
approximation. The path operator of the CPA effective medium is

\begin{equation}
\mathbf{\tilde{g}}_{LL'}({\mathbf k},z) =
\frac{1}{\mathbf{S}_{LL'}({\mathbf k},z) - \mathbf{\tilde{D}}_{LL'}(z)} ,
\end{equation}
where $\mathbf{\tilde{g}}_{LL'}({\mathbf k},z)$ and
$\mathbf{\tilde{D}}_{LL'}(z)$ are the path operator and potential
function of the CPA effective medium, respectively. The latter is
in general a non-diagonal matrix in the angular moment representation.

The CPA effective medium path operator and potential function are found
from the set of CPA equations\cite{CPA} but in the spinor representation:

\begin{eqnarray} \label{eq:CPA}
&&\mathbf{\widetilde{g}}_{LL'}(z) =
\int d {\mathbf k} \mathbf{\tilde{g}}_{LL'}({\mathbf k},z)
\equiv  \mathbf{\widetilde{g}} \\
&&\mathbf{g}^{\alpha} = \mathbf{\widetilde{g}} + \mathbf{\widetilde{g}}
\left[\mathbf{D}^{\alpha} - \mathbf{\tilde{D}}\right]
\mathbf{g}^{\alpha} \\
&&\mathbf{\widetilde{g}} = \sum_{\alpha} c^{\alpha} \mathbf{g}^{\alpha} .
\end{eqnarray}
Here, $c^{\alpha}$, $\mathbf{D}^{\alpha}$, and $\mathbf{g}^{\alpha}$
are the concentration, potential function and
on-site path operator of alloy component $\alpha$ in the
CPA effective medium. Both, $\mathbf{D}^{\alpha}$ and $\mathbf{g}^{\alpha}$,
are digonal in the spinor representation and therefore the rest of the
formalism for the self-consistent and total energy calculations is similar
to the one for the collinear EMTO-CPA method.\cite{vitos01} 

\section{Details of the spin-wave method calculations}
\label{SWM}

The total energy of a system in magnetic configuration $\alpha$, specified
by its long-range order parameter, $s_0^{\alpha}$, and spin-spin correlation
functions, $\xi^{\rm SRO}_{\alpha}(\mathbf{R})$ or
$\xi^{\rm SRO}_{\alpha}(\mathbf{q})$, is given by (\ref{eq:E_fm}).
In the actual spin-wave calculations, when
the Monkohrst-Pack\cite{monkhorst72} q-point grid is used, the total
energy of the system in magnetic state $\alpha$  for a particular
volume per atom, $\Omega$, is

\begin{eqnarray} \label{eq:E_fm_sum}
E^{\alpha}(\Omega) &=& s_0^{\alpha} E(\mathbf{q}=0,\Omega) + \\ \nonumber
&& \sum_{i} w_i E(\mathbf{q}_i,\Omega)\left[1 -s_0^{\alpha} + 
\xi^{\rm SRO}_{\alpha}(\mathbf{q}_i)\right] ,
\end{eqnarray}
where $\mathbf{q}_i$ and $w_i$ are the q-points of the grid and their
weights. $E(\mathbf{q}_i,\Omega)$ is the total energy of a system
for a planar spin-spiral ($\theta=\pi/2$) with wave vector $\mathbf{q}_i$
and volume per atom $\Omega$.  $E(\mathbf{q}=0,\Omega)$ is the energy of
the ferromagnetic state (point $\mathbf{q}=0$ can be in the grid or
calculated separately). 

Thus, the set of the total energies $E(\mathbf{q}_i,\Omega)$ 
on the q-point grid uniquely
defines the total energy of the system in any given magnetic
configuration $\alpha$, $E^{\alpha}(\Omega)$. The accuracy of such a
representation of a particular magnetic state can be always checked for
the given grid of q-points by
calculating  $\xi^{\rm SRO}_{\alpha}(\mathbf{R})$ from 
$\xi^{\rm SRO}_{\alpha}(\mathbf{q}_i)$.\cite{ruban12} Although,
obviously, there always be a problem with 
$\xi^{\rm SRO}_{\alpha}(\mathbf{R})$ for $R \to \infty$  when
the finite grid is used, in most cases it cannot affect the
result for the total energy due to the relatively short-range
character of the magnetic exchange interactions, which is in
particular the case of Fe-Ni alloys, where the the 7$\times$7$\times$7
Monkhorst-Pack grid provides quite accurate total energies
for the whole range of magnetic states.

\begin{acknowledgments}
The author acknowledges the support of the Swedish Research Council
(VR project 2015-05538), a European Research Council grant,
the VINNEX center Hero-m, financed by the  Swedish Governmental
Agency for Innovation Systems (VINNOVA), Swedish industry, and
the Royal Institute of Technology (KTH). Calculations were done
using NSC (Link\"oping) and PDC (Stockholm) resources provided
by the Swedish National Infrastructure for Computing (SNIC).
The support from the Austrian federal government
(in particular from Bundesministerium fŸr Verkehr, Innovation und
Technologie and Bundesministerium fŸr Wirtschaft, Familie und Jugend)
represented by …sterreichische Forschungsfšrderungsgesellschaft mbH
and the Styrian and the Tyrolean provincial government, represented
by Steirische Wirtschaftsfšrderungsgesellschaft mbH and Standortagentur
Tirol, within the framework of the COMET Funding Program is
also gratefully acknowledged.
\end{acknowledgments}


\begin{thebibliography}{99}
\bibitem{at-wt} In the Materials Science and Engineering literature weight 
\% are usually used without mentioning. Close to the Invar composition,
wt. \% of Ni is about 1 \% higher than atomic \%.
\bibitem{superInvar} For instance, the much lower thermal expansion
coefficient at room temperature can be obtained using Co additions,
like in the case of super Invar alloy, Fe$_{63}$Ni$_{31}$Co$_{6}$.
\bibitem{scell} The choice of the alloy composition, Fe$_{62.5}$Ni$_{37.5}$,
was dictated by the optimization of the supercell size under
constrain having the first eight pair correlation functions as small
as possible, i.e. to represent best a random alloy up to the 
8th coordination shell. Let us note, that this alloy composition
is still in the Invar region.
\bibitem{j_xc_LDA} 
Let us note, that the GGA also increases the magnitude of the magnetic
moment compared to the LDA, and the GGA exchange interaction will
produce higher Curie temperatures by approximately 150 K (see,
for instance, Ref. \onlinecite{ruban04}).
\end{thebibliography}
\end{document}